\documentstyle[aaspp4]{article}

\newcommand{\Bnu}{\mbox{$B_\nu$}}

\newcommand{\Fnu}{\mbox{$F_\nu$}}

\newcommand{\Inu}{\mbox{$I_\nu$}}

\newcommand{\kmsec}{\mbox{$\rm{km \; s^{-1}}$}}

\newcommand{\beq}{\begin{equation}}
\newcommand{\eeq}{\end{equation}}
\newcommand{\epsilono}{\epsilon_0}
\def\lae{\mathrel{<\kern-1.0em\lower0.9ex\hbox{$\sim$}}}
\newcommand{\gae}{\mathrel{>\kern-1.0em\lower0.9ex\hbox{$\sim$}}}

\newcommand{\mH}{\mbox{$m_{\rm H}$}}

\newcommand{\mdot}{\mbox{$\dot{M}$}}

\newcommand{\nuz}{\mbox{$\nu_{\rm z}$}}

\newcommand{\rvec}{\mbox{$\bf r$}}

\newcommand{\Snu}{\mbox{$S_\nu$}}

\newcommand{\taul}{\mbox{$\tau_l$}}

\newcommand{\vvec}{\mbox{$\bf v$}}
\newcommand{\vrad}{\mbox{$v_{\rm r}$}}
\newcommand{\vinf}{\mbox{$v_\infty$}}
\newcommand{\vz}{\mbox{$v_{\rm z}$}}

\newcommand{\zhat}{\mbox{$\bf \hat{z}$}}

\def \etal{et~al.\/}

\begin{document}

\title{Profile Shapes for Optically Thick X-ray Emission Lines
from Stellar Winds}

\author{R.\ Ignace\altaffilmark{1,}\altaffilmark{2} and K.~G.~Gayley\altaffilmark{2}}
 
\altaffiltext{1}{
        Email:  ri@astro.physics.uiowa.edu }
 
\altaffiltext{2}{
	203 Van Allen Hall, 
	Department of Physics and Astronomy,
	University of Iowa, 
	Iowa City, IA 52242 USA }

\begin{abstract}

We consider the consequences of appreciable 
line optical depth for the profile
shape of X-ray emission lines formed in stellar winds.  The hot
gas is thought to arise in distributed wind shocks, and the line
formation is predominantly via collisional excitation followed by
radiative decay.  Such lines are often modelled as optically thin,
but the theory has difficulty matching resolved X-ray line
profiles.  We suggest that for strong lines of abundant metals,
newly created photons may undergo resonance scattering, modifying
the emergent profile.  Using Sobolev theory in a spherically
symmetric wind, we show that thick-line resonance scattering leads
to emission profiles that still have blueshifted centroids like
the thin lines, but which are considerably less asymmetric in
appearance.  We focus on winds in the constant-expansion domain,
and derive an analytic form for the profile shape in the limit of
large line and photoabsorptive optical depths.  In this limit the
emission profile reduces to a universal shape and has a centroid
shift of $-0.24\vinf$ with a HWHM of $0.63\vinf$.  Using published
data for {\it Chandra} observations of five emission lines from
the O~star $\zeta$~Pup, we find that the observed HWHMs are somewhat
smaller than predicted by our theory; however, the centroid shifts
of all five lines are consistent with our theoretical result.
These optical-depth effects can potentially explain the more nearly
symmetric emission lines observed in $\zeta$~Ori, $\theta^1$~Ori~C,
and $\delta$~Ori by {\it Chandra}, although an alternative explanation
is required to account for the unshifted peak line emission.  We
also consider enhanced reabsorption by continuous opacity as line
photons multiply scatter within an optically thick line, and find
for lines with optical depths of a few that such reabsorption can
further reduce the line asymmetry.  It also reduces the line
equivalent width, but probably not enough to alleviate the problem
of sub-solar metallicities inferred from O~star X-ray spectra by
{\it ASCA}, unless the width of the resonance regions are superthermally
enhanced.


\end{abstract}

\twocolumn

\section{INTRODUCTION}

Observations by the {\it Chandra} and {\it XMM-Newton} telescopes
are providing first-ever resolved emission line profiles from
early-type stellar winds in the X-ray band (the O~star $\zeta$ Pup
by Kahn \etal\ 2000 and Cassinelli \etal\ 2001; the O~star
$\theta^1$~Ori~C by Schulz \etal\ 2001; the O~star $\zeta$ Ori by
Waldron \& Cassinelli 2001; and the O~star $\delta$~Ori by Miller
\etal\ 2001).  Already, many early-type stars were known to be X-ray sources
from observations by {\it EINSTEIN} and {\it ROSAT} (Harnden
\etal\ 1979; Seward \etal\ 1979; Seward \& Chlebowski 1982; Pollock
1987; Pollock, Haberl, \& Corcoron 1995; Kudritzki \etal\ 1996;
Bergh\"{o}fer \etal\ 1997).  Although early attempts sought to
explain the X-ray emission via stellar coronae (e.g., Cassinelli
\& Olson 1979; Waldron 1984), the currently favored model is based
on considerations of stellar-wind shocks (e.g., Lucy \& White 1980;
Lucy 1982; Owocki, Castor, \& Rybicki 1988; Feldmeier, Puls, \&
Pauldrach 1997).  The highly successful line-driven wind theory
for explaining hot star winds (Lucy \& Solomon 1970; Castor, Abbott,
\& Klein 1975; Friend \& Abbott 1986; Puls, Pauldrach, \& Kudritzki
1986) is well suited to generating instabilities leading to shock
structures distributed throughout the wind flow and giving rise to
the production of hot gas.

The confrontation of theory with data has led to mixed results.  For
example, Feldmeier \etal\ (1997) conducted extensive
modelling of wind shock X-ray production using time-dependent
hydrodynamical simulations.  Their sophisticated treatment includes a
voluminous line list, but is limited to spherical symmetry, and only
models for O~stars have been investigated.  Within this context, their
efforts reveal that a single high-density shell will tend to dominate
the observed X-ray emission.  Such a feature should lead to significant
variability which has not been observed (e.g., see Bergh\"{o}fer
\& Schmitt 1995).  Phenomenologically, perhaps their results apply in a
``piecewise spherical'' sense, whereby these X-ray bright events no
longer occur in a single global shell, but as numerous smaller events
occuring at different times.  In this way, the variability can be
suppressed (e.g., Oskinova \etal\ 2001).

Aside from the issues of global X-ray production and variability, the
resolved emission line profiles in four O~stars ($\zeta$~Pup,
$\zeta$~Ori, $\theta^1$ Ori~C, and $\delta$~Ori) appear to be truly
problematic for the current theory.  The X-ray emission is expected
to be dominated by a spectrum of optically thin lines.  In addition
to the intervening ISM, the stellar wind itself can significantly
contribute to the attenuation of shock generated X-rays.  Consequently,
the observed emission profiles are expected to be asymmetric, with
peak line emission appearing blueshifted of line center, with the
redshifted emission being more suppressed because there is a larger
column depth of attenuation to the receding hemisphere of the wind
(MacFarlane \etal\ 1991; Ignace 2001; Owocki \& Cohen 2001).
However, the observed line profiles for $\zeta$~Ori, $\theta^1$
Ori~C, and $\delta$~Ori appear to be quite symmetric.  The emission
lines for $\zeta$~Pup are far more consistent with the expectations
of distributed wind shocks, but even for this star, the line
morphologies may not be consistent with the current theory.  So
either the underlying wind model, or simply the assumptions used
in deriving the emission-line profile shapes, must be modified.
It is this latter case that is the focus of this work, with particular
attention given to $\zeta$~Pup since it appears to best exemplify
current standard wind-shock expectations.

Raymond \& Smith (1977) made a major contribution to interpreting
the X-ray emission from hot astrophysical plasmas by providing a
cooling function that, among other processes, accounts for the
emission by optically thin lines.  Their results, along with better
line data, have been used in the modelling efforts of numerous studies,
including the topic at hand of stellar winds.  However, some have
questioned the assumption of effectively thin emission among the
stronger wind lines in the X-ray band (e.g., Hillier \etal\ 1993;
Kitamoto \& Mukai 1996; Kitamoto \etal\ 2000).  As will be discussed
in \S 2, we find that it is possible for some lines to be optically
thick.  In \S 3, the formation of optically thick line profiles in
distributed X-ray emitting zones for a stellar wind is considered
using standard Sobolev theory.  In
particular, focus is given to the case of constant spherical
expansion appropriate for relatively dense winds in which the X-ray
emission escapes only from several radii out in the flow.  This
case is germane to the O~stars which have been observed with {\it
Chandra} and {\it XMM}.  Optically thick emission not only alters
the profile diagnostics, it presents the possibility for X-ray
photon destruction, which would invalidate the Raymond \& Smith
assumption for those lines.  In \S 4, we consider the role of photon
absorption in X-ray bound-free continua {\it within} the resonance
zones in which the photons are created.  A discussion of our results
in relation to published observations is given in \S 5.

\section{THE OPTICAL DEPTH OF X-RAY LINES}

Should one expect wind emission lines produced in the X-ray band
to be optically thick?  As already stated, the primary process by
which X-ray line photons are created is collisional excitation
followed by radiative decay.  For minority species, radiative
recombination into excited states may also contribute, but for our
purposes this may be viewed as essentially a collisional process
as well.  The key assumption is that the photon creation rate per
volume scales with the square of the density.  To estimate the
optical depth, we consider a slab of X-ray emitting shocked gas.
For a slab of initial temperature $T_X$, constant electron number
density $n_{\rm e}$, and volume cooling rate $\Lambda(T_X)$, the
cooling time will be given roughly by

\begin{equation}
t_{\rm c} \approx \frac{5}{2}\,\frac{k\,\Delta T}{n_{\rm e}\,\Lambda(T_X)}.
\end{equation}

\noindent The $\Delta T$ is to represent the range of temperatures
over which a particular ion state $i$ of some atomic species $j$
exists in the slab for the purpose of estimating the line optical
depth.  The expression thus describes the cooling time for a
particular ion species from $T_X+0.5\Delta T$ to $T_X-0.5\Delta
T$.

The line optical depth is given roughly as

\begin{equation}
\tau \approx n_1\, \sigma_l\,l_{\rm c}\,\Delta \nu^{-1}, 
\end{equation}

\noindent where $n_1$ is the population number density of the ground
state for the transition of interest, $\sigma_l$ is the line integrated
optical depth as given by

\begin{equation}
\sigma_l \approx \frac{\pi\,e^2}{m_{\rm e}\,c} \, f_{12},
\end{equation}

\noindent which applies for a two-level atom that is strongly 
NLTE, $l_{\rm c}$ is the cooling length as given by

\begin{equation}
l_{\rm c} = V\,t_{\rm c},
\end{equation}

\noindent for $V$ the speed of the postshock material, and $\Delta
\nu = v_D/\lambda$ is the frequency width of the line opacity for
Doppler broadening $v_D$ and line wavelength $\lambda$.

For simplicity the Doppler broadening is just taken as the thermal
speed of the hot gas, although turbulent motions could be important
and somewhat larger.  For NLTE, the population number density of
the ground state is roughly equal to the entire number density for
that ion $n_{\rm i}$.  Taking this to be the case, $n_{\rm i}$ can
be re-expressed in terms of the ionization fraction $q_{\rm i}$ of
the ion to all ions of the species and the relative abundance ${\cal
A}_{\rm j}$ of the species to all nuclei $n_{\rm N}$, giving $n_{\rm
i} \approx q_{\rm i}\,{\cal A}_{\rm j}\,n_{\rm N}$.  Thus the
optical depth of the line formed in just this one slab becomes

\begin{equation}
\tau \approx \frac{5\,k\,\Delta T\,q_{\rm i}\,{\cal A}_{\rm j}\,n_{\rm N}\,
	\sigma_l\,\lambda\,V}{2\,n_{\rm e}\,\Lambda(T_X)\,v_{\rm th}} .
\end{equation}

We estimate the optical depth for \ion{O}{8} Ly$\alpha$ at 19.0
\AA\ as an example of a strong line in a gas with a solar composition
of oxygen (${\cal A}_j = 9\times 10^{-4}$).  From Sutherland \&
Dopita (1993), the ionization fraction of \ion{O}{8} peaks at around
2~million~K with a value of $q_{\rm i} \approx 0.5$, which
predominantly exists over a temperature range of $\Delta T \approx
2$~million~K.  The volume cooling rate is about $6\times
10^{-23}$~erg~s$^{-1}$~cm$^3$.  The necessary shock jump is
500~km~s$^{-1}$, and the thermal speed for hot oxygen is 45~km~s$^{-1}$.
Hence the optical depth for this one slab becomes $\tau \approx
10^2$.  This demonstrates that it is entirely reasonable that the
strong lines of metals with ${\cal A}_j \gtrsim 10^{-4}$, which
includes many of the lines observed in hot star X-ray spectra, 
could be optically thick.

Having shown that optical thickness may be relevant for strong
resonance lines, what should we expect for its influence on the
emission profile shape?  The observed profiles for $\zeta$~Pup (Kahn
\etal\ 2001; Cassinelli \etal\ 2001) are asymmetric with blueshifted
peak emission.  As will be discussed further at a later point,
photoabsorption by the ambient wind component generally accounts
for this asymmetry.  However, the effect of line optical depth is
to alter the primary direction of photon escape.  For example, the
emissivity for a line that is thin is isotropic, and so there is
no preferred escape direction for photons.  For a radially narrow
but optically thick slab as considered above, photons escape
primarily normal to the slab face, since the optical depth in that
direction is least.  For a large number of slabs at a fixed radius
and uniformly distributed in solid angle, and ignoring the wind
photoabsorption, those slabs that are fore and aft will tend to be
the brightest based on this argument.  Now accounting for the
asymmetric photoabsorption across the line frequencies, this suggests
that peak blueshifted emission should occur near the maximum Doppler
shift.  Instead, the observed emission profiles tend to have peak
emission intermediate between line center and the extreme wing.

One other possibility is that photons escape preferentially in the
{\it lateral} direction.  The motivation here is that in outflows,
velocity gradients can lead to just such escape avenues.  In this
case, consider numerous discrete slabs distributed across a spherical
shell.  All else being equal, the brightest slabs will be those
located near the plane of the sky owing to the lateral escape of
radiation, and corresponding to frequencies around line center,
which is more in accord with observations.  

However, we caution that the Sobolev approximation applies when
the Sobolev length $l_{\rm sob}$ is significantly smaller than the
cooling length $l_{\rm c}$ (i.e., the extent of the hot gas region).
A rough expression for a ratio of the two is

\begin{equation}
\frac{l_{\rm c}}{l_{\rm sob}} \approx \left(\frac{10^{10}}{n_{\rm e}}\right)\,
	\left(\frac{v_{\rm sh}}{v_{\rm th}}\right)\,\left(
	\frac{R_\odot}{R_*}\right),
\end{equation}

\noindent where $n_{\rm e}$ is in cm$^{-3}$ and $v_{\rm sh}$ is
the speed of the shocked gas in the observer's frame.  We will
consider a spherically symmetric expanding wind with ambient number
density

\begin{equation} 
n_{\rm w}(r) = \frac{\dot{M}}{4\pi\,\mu\,m_H\,r^2\,\vrad(r)},
\end{equation}

\noindent where $\dot{M}$ is the mass-loss rate, $\vrad$ is the
radial wind speed, and $\mu$ is the mean molecular weight per
particle. The electron number density will naturally scale with
the wind density.  For typical hot stars, $n_{\rm w}$ ranges from
$10^{10}$--$10^{12}$ cm$^{-3}$ at the wind base and drops to
$10^7$--$10^9$ by about $3R_*$, thereafter decreasing as $r^{-2}$.
The Sobolev approximation should thus be valid at most radii, except
possibly in the densest regions of the inner wind acceleration
where $\vrad$ is relatively small.  But even at small radii, the
Sobolev approach may be adequate if the shocked gas is of low
density, as would be the case if the emission arises in reverse
shocks.  Alternatively, if $R_* \gg l_{\rm sob} \gg l_{\rm c}$,
then there exist numerous pockets of hot gas within a Sobolev
length.  These would likely be optically thin, so treating them as
statistically homogeneous allows us to recover the Sobolev
approximation.  

However, to say that the photons escape laterally implies that the
lateral optical depth is less than the radial optical depth.  So
the effect of the velocity gradient carves the slab up into numerous
radiatively independent and elongated ``columns''.  The short length
of these columns is the Sobolev length, and the long length is the
slab thickness.  Given that the optical depth through the slab's
thickness was shown to be $\approx 10^2$, the columns can still be
thick in the lateral direction for $l_{\rm c}/ l_{\rm sob}$ ranging
from 10--100.  This range is entirely reasonable given the preceding
estimates, and so we next apply the Sobolev theory to deriving
emission profile shapes to model the effects of lateral photon
escape.

\section{OPTICALLY THICK X-RAY EMISSION PROFILE SHAPES}

\subsection{Formalism for the Line Formation}
	\label{sub:linform}

For an envelope in bulk motion such that the flow speed greatly
exceeds the thermal broadening, the locus of points contributing
to the emission at any particular frequency in the line-profile is
confined to an ``isovelocity zone'' (c.f., Mihalas 1978).  These
zones are determined by the Doppler-shift formula, namely

\begin{equation} 
\nuz = \nu_0\,\left(1-\frac{\vz}{c}\right),
	\label{eq:doppler} 
\end{equation}

\noindent where $\nuz$ is the Doppler-shifted frequency and $\vz
= - \vvec(\rvec)\cdot\zhat$ is the projection of the flow speed
onto the line of sight along $z$.  The familiar Sobolev theory for
line-profile calculation in moving media reduces the radiation
transfer in the envelope to radiation transfer within isovelocity
zones (i.e., the transfer between different points in
the envelope decouples if their relative velocity shift exceeds
the local line broadening).  Thus, the emergent line intensity along
a given ray through the extended stellar envelope is given by

\begin{equation}
\Inu = \Snu(r)\,\left[1-e^{-\tau_l(r,\mu)}\right]\,e^{-\tau_{\rm c}
	(r,\mu)},	\label{eq:inten}
\end{equation}

\noindent where \Snu\ is the source function, $\tau_l$ is the
Sobolev optical depth of the line where the ray intersects the
isovelocity zone, and $\tau_{\rm c}$ is the intervening absorptive
optical depth.  

Allowing for line emission from resonance line scattering and
collisional de-excitation, and defining $\epsilon$ as the ratio of
the collisional de-excitation rate (related to the rate of collisional
excitation) to that of spontaneous decay, the source function is
given by

\begin{equation} 
\Snu = \frac{\beta_{\rm c}\,I_{\nu,*}+\epsilon\,\Bnu}{\beta+\epsilon\,
	(1-\beta)}.
\end{equation}

\noindent The two parameters $\beta_{\rm c}$ and $\beta$ are
respectively the penetration and escape probabilities, defined as

\begin{equation}
\beta_{\rm c} = \frac{1}{4\pi}\,\int_{\Omega_*}\,
        \frac{1-e^{-\tau_l}}{\tau_l}\,d\Omega,
\end{equation}
 
\noindent and
 
\begin{equation}
\beta = \frac{1}{4\pi}\,\int_{4\pi}\,
        \frac{1-e^{-\tau_l}}{\tau_l}\,d\Omega.	\label{eq:beta}
\end{equation}

Adopting standard cylindrical coordinates $(p,\alpha,z)$, with the
observer at $+\infty$, the line optical depth is given by

\begin{equation}
\tau_l(p,z) = \int_{z}^\infty\, \sigma_l\,n_1\,\delta(\nu-\nu_{\rm z})\,
        dz'= \frac{\sigma_l\,n_{\rm k}(r)\,\lambda_0}{| d\vz/dz |_{\rm p} },
	\label{eq:tausob}
\end{equation}
 
\noindent where $\sigma_l$ is the frequency-integrated line
cross-section, and the denominator on the right-hand side is the
line-of-sight velocity gradient.  In the case of a spherical flow,
the line-of-sight velocity gradient is

\begin{equation}
\frac{d\vz}{dz} = \mu^2\,\frac{dv}{dr} + (1-\mu^2)\,\frac{v}{r},
	\label{eq:vgrad}
\end{equation}

\noindent for $\mu=\cos \theta$ the direction cosine from the radial to
the observer.  

For the case of X-ray emission, we assume that the star is X-ray
faint (i.e., no basal corona) such that $I_{\nu,*}=0$ at X-ray
energies.  Also, the destruction probability $\epsilon$ depends on
density, and so drops with radius.  To make this explicit, we parameterize

\begin{equation}
\epsilon = \epsilon_0\, \rho/\rho_0  ,
\end{equation}

\noindent where $\rho_0$ and $\epsilono$ are fiducial values with
$\epsilon_0 \ll 1$ being typical.  Under these conditions, the
source function becomes

\begin{equation}
\Snu \approx \frac{\epsilon}{\beta}\, \Bnu.
\end{equation}

\noindent To compute the total emergent flux at a given Doppler
shift in the line profile, we integrate over all intensity beams
to obtain

\begin{equation} 
\Fnu(\vz) = \frac{1}{D^2}\,\int_{v_{\rm
	z}}\, \Inu(p,z)\,p\,dp\,d\alpha ,
	\label{eq:flux} 
\end{equation}

\noindent where $D$ is the distance from the Earth.  Self-absorption by
the hot plasma is ignored.  The attenuation is therefore entirely from
the cool wind component intervening between the observer and the point
of emission.  (For actual observations, one must also correct for the
energy-dependent interstellar extinction, but we ignore its effects
here since it does not alter the profile shape over the narrow extent of
the line width.) 

The wind optical depth is given by

\begin{equation} 
\tau_{\rm c}(p,z) = \int_z^{\infty}\,\kappa_{\rm c}\,\rho_{\rm w}(p,z)\,dz
\end{equation}

\noindent with opacity $\kappa_{\rm c}$ and density $\rho_{\rm w}$.  The
dominant opacity at the X-ray energies is photo-absorption by K-shell
electrons.  This opacity is calculated from

\begin{equation} 
\kappa_{\rm c}(E) = \frac{1}{\mu_{\rm N} \mH}\,\sum_{\rm j}
	\, \frac{n_{\rm j}}{n_{\rm N}} \, \sigma_{\rm j} (E).  
\end{equation}

\noindent The opacity is a summation over cross-sections presented by
different atomic species $j$ and weighted by the relative abundance
$n_{\rm j}/n_{\rm N}$, for $n_{\rm N}$ the number density of nuclei.

The line source function and a prescription for the wind attenuation
are all the required pieces necessary to compute X-ray emission
line profiles from stellar winds.  The above expressions are applied
in the next section for the specific case of a constant expansion flow.

\subsection{The Constant Expansion Case}
	\label{sub:const} 

\subsubsection{Justification of Assumption}

The constant expansion wind is the most relevant case to consider for
the problem at hand, since the continuum absorption by the winds
of most O~stars are expected to be optically thick over much of
the X-ray band.  The implication is that the observed X-rays emerge
from the wind at depths of a few stellar radii or more.  For the
O~star winds, the standard $\beta$-law velocity prescription for
the radial distribution of wind speeds begins to asymptote around
1 to 1.5$R_*$ above the wind base for the commonly used value of
$\beta\approx 1$.  Consequently for many lines, it is sufficient
to model the emission profiles as if the entire wind were undergoing
constant expansion.  This case is also convenient because the
expressions for the radiative transfer become analytic and provide
interesting insight for better understanding the resulting profile
shapes.  However, there is at least one observational example of
a line that seems to originate from deeper levels in the flow where
the constant expansion assumption no longer applies (e.g., He-like
\ion{S}{15} observed in $\zeta$~Pup by Cassinelli \etal\ 2001).
In \S\ref{sec:disc}, we comment on the effect of the wind acceleration
zone on the profile shape.

We begin by motivating the constant expansion assumption on more
quantitative grounds.  Owocki \& Cohen (1999; for OB stars) and
Ignace \& Oskinova (1999; for WR~stars) presented a scaling analysis
for the X-ray emission from hot star winds based on an ``exospheric''
approximation.  The observed X-ray emission (assumed to arise from
an optically thin hot plasma) emerges only from radii exterior to
the optical depth unity surface of radius $r_1$, with X-rays at
smaller radii assumed to be completely attenuated.  The extent of
this convenient scale $r_1$ derives from the photoabsorption by
the wind along the line-of-sight.  It is determined by the condition
that

\begin{equation} 
\tau_{\rm c} \approx 1 = \int_{r_1}^\infty \, \kappa_{\rm c}\,\rho_{\rm w}\,dr.
	\label{eq:tauc1}
\end{equation}

\noindent Assuming that $v(r)=\vinf$ allows
for an analytic evaluation of this integral, yielding

\begin{equation} 
r_1 (E) = \frac{\dot{M}}{4\pi v_\infty}\,\kappa_{\rm c} (E).
\end{equation}

\noindent This may seem circular, since we apply the constant
expansion assumption here to justify that constant expansion should
be adequate for the emission line calculation.  But in fact, the
integral of equation~(\ref{eq:tauc1}) minimizes the value of $r_1$
in the case of $v_{\rm r}=\vinf$, hence this is the conservative
approach to showing that $r_1 \gtrsim 2R_*$ is valid over much of
the X-ray band.

\begin{figure}
\plotone{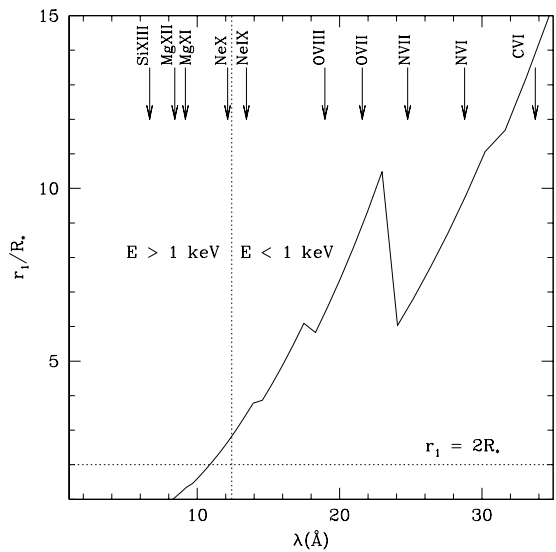}
\caption[]{A plot of the optical depth unity radius from the wind
photoabsorptive opacity versus wavelength in the X-ray band.  Typical
O~star and wind parameters are assumed.  The horizontal dotted line
indicates $r_1=R_*$, and the vertical line gives the location of
$E=1$~keV, corresponding to $\lambda = 12.4\AA$.  
The location of strong emission lines are noted.

\label{fig:f1}}
\end{figure}

Using typical O~star and wind parameters as in \S 2, a plot of
$r_1$ versus wavelength is shown in Figure~\ref{fig:f1}.  The
``jagged'' features correspond to prominent photoabsorptive edges.
Observed strong lines of H and He-like ions in stellar winds are
generally located between 5 and 30~\AA, as indicated at top.  Note
that 1~keV corresponds to about 12.4~\AA\ (indicated by the vertical
line in the plot), and so harder X-ray energies are at shorter
$\lambda$, and vice versa.  A horizontal line is provided where
$r_1=2R_*$, beyond which the wind is essentially in constant
expansion for O~star winds.  Many of the strong lines of interest
occur at wavelengths for which $r_1 \gtrsim 2R_*$, where the
approximation that $v_{\rm r} \approx \vinf$ will hold.  A similar
figure is provided by Cassinelli \etal\ (2001) in their discussion
of {\it Chandra} observations of $\zeta$~Pup.  They even tabulate
$r_1$ values for several lines in their Table~2.  It happens that
$\zeta$~Pup has a slightly higher \mdot\ and lower \vinf\ than we
have used, so that even for \ion{Mg}{12} at 8.42~\AA, Cassinelli
\etal\ find $r_1 = 2.8R_*$.

\subsubsection{Applying the Constant Expansion Assumption}

Concluding that constant expansion is a reasonable assumption, we
now consider again the line formation process under this condition.
>From equation~(\ref{eq:vgrad}), the line-of-sight velocity gradient
reduces to

\begin{eqnarray}
\label{taulgen}
\tau_l & = & \frac{\sigma_l\,n_X(r)\,\lambda_0}{(\vinf/r)\,(1-\mu^2)}  \\
 & = & T\,\left(\frac{R_*}{r}\right)\,(1-\mu^2)^{-1},
	\label{eq:tauconst}
\end{eqnarray}

\noindent where an optical depth scale $T$ has been introduced.
This scale is the ``integrated optical depth'' of the line along
the line-of-sight (Groenewegen, Lamers, \& Pauldrach 1989).  The
parameter $T$ can be related to the column density of scatterers
$N_1$ as follows,

\begin{eqnarray}
N_1 & = & \int_R^\infty \, n_1 \, dr \\
 & = & \int_0^{v_\infty}\, \frac{\tau_l}{\sigma_l\,\lambda_0}\,d\vrad.
\end{eqnarray}

\noindent Defining $T = \int_0^1\,\tau_l\,dw$ for $w=\vrad/\vinf$,
we obtain

\begin{equation}
T = \frac{\sigma_l\,\lambda_0}{\vinf}\,N_1.
	\label{eq:Tdefn}
\end{equation}

Unfortunately in the constant expansion approximation, the Sobolev
optical depth of equation~(\ref{eq:tauconst}) contains a pole at
$\mu= \pm 1$, corresponding to the extreme blue and red shifts.
Physically of course, there is no real singularity problem.  Because
the wind is in constant expansion, there is no radial velocity
gradient, hence the actual optical depth in the vicinity of $\mu
= \pm 1$ is just $(\vinf/v_{\rm th})\,T$, for $v_{\rm th}$ the
thermal broadening, because the radiative transfer corresponds to
the case of a slab moving at constant speed $\vinf$.  Since
$\vinf/v_{\rm th}\gtrsim 10$ can be expected, values of $T$ exceeding
a few will ensure that $\tau_l$ for $\mu \approx \pm 1$ will be
large enough that the exponential factors in the Sobolev intensity
of equation~(\ref{eq:inten}) and the escape probability of
equation~(\ref{eq:beta}) are effectively zero, hence the ``pole''
problem will not be of significant concern.  There are two subsequent
implications for the constant expansion wind.  (a) Instead of
the line-of-sight integrated optical depth, $T$ actually
corresponds to $\tau_l$ at $\mu=0$ and is thus a lateral optical
depth scale.  This is reasonable since for constant expansion, the
line-of-sight velocity gradient arises only from the geometrical
divergence of the spherical wind outflow.  This scale is also the
minimal Sobolev optical depth of $\tau_l(\mu)$ for
any fixed $r$.  (b) The radial optical depth can
be large, and so there is the possibility for effective photon
trapping and substantial reabsorption by continuum opacity.  This
effect will be considered in \S4.

Given that the thin line case has already been explicitly treated, we
choose to focus on the thick line case.  The major simplification that
results comes in the escape probability.  Recall that $\beta$ varies
between 0 and 1.  For thin lines, $\beta$ approaches the value of unity,
meaning that photons scatter isotropically.  For thick lines, $\beta$
decreases to small values.  Taking $\tau_l \gtrsim 1$, the escape
probability reduces to

\begin{eqnarray}
\label{betaapp}
\beta & \approx & \frac{1}{2}\,\int_{-1}^{+1}\, \left(\frac{1}{T}\right)\,
	\left(\frac{r}{R_*}\right)\,(1-\mu^2)\,d\mu \\
 & = & \frac{2}{3\,T}\,\left( \frac{r}{R_*} \right) .	\label{eq:betaconst}
\end{eqnarray}

\begin{figure}
\plotone{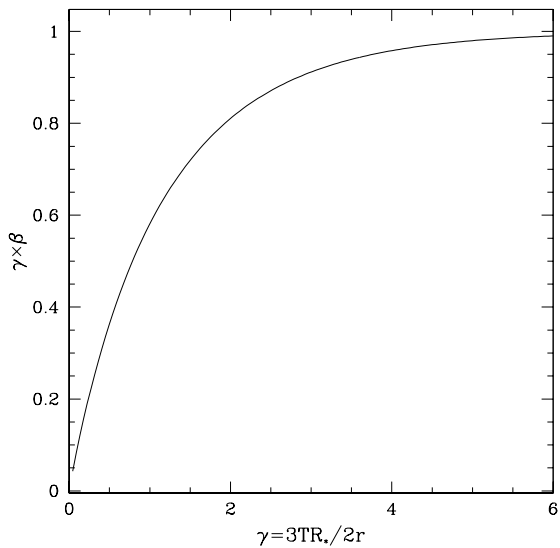}
\caption[]{A plot of how rapidly the line escape probability approaches
the asymptotic limit of $\beta = \gamma^{-1} = 2r/3TR_*$.  The vertical
is shown as the product $\gamma\beta$.  Clearly $\beta$ is nearing
its asymptotic form at $\gamma \approx 2$.

\label{fig:f2}}
\end{figure}

\noindent Clearly at large radius where $\tau_l$ drops below unity, the
approximation breaks down.  Defining $\gamma = 3TR_*/2r$,
Figure~\ref{fig:f2} shows a plot of $\gamma\cdot\beta$, indicating that
this product begins to plateau at unity around $\gamma$ of 2.  Since
the constant expansion approximation applies only for $r/R_*$ exceeding
about 2--3, values of $T\gtrsim 10$ are required for lines to show
interesting optical depth effects and for
expression~(\ref{eq:betaconst}) to apply.

Continuing in the assumption that the line is optically thick, and
noting that for constant expansion $\epsilon = \epsilon_0\,(R_*/r)^2$,
the source function reduces to the form

\begin{equation}
\label{snur}
\Snu(r) \approx \frac{3}{2}\,T\,\epsilono\,\Bnu \,\left(\frac{R_*}{r}
	\right)^3.
\end{equation}

\noindent The emergent intensity from a given point $(r,\mu)$ is given by

\begin{equation}
\Inu \approx \Snu(r)\, e^{-\tau_{\rm c}(r,\mu)},
\end{equation}

\noindent where the continuum optical depth can be derived analytically
(e.g., MacFarlane \etal\ 1991; Ignace 2001) as

\begin{eqnarray}
\tau_{\rm c}(r,\mu) & = & \tau_0 \, \left(\frac{R_*}{r}\right)\,\left(\frac{\theta}{\sin\theta}\right) \\ \label{eq:tauc}
 & = & \tau_0(E) \, \left(\frac{R_*}{r}\right)\,\frac{\cos^{-1} \mu}{\sqrt{1-\mu^2}},
\end{eqnarray}

\noindent for $\tau_0(E) = \kappa_{\rm c} (E) \mdot/4\pi R_*\vinf$, the
total continuum optical depth to the star along the line-of-sight.

To calculate the emission profile then requires an integration over
the isovelocity zone corresponding to each frequency in the profile.  
For constant expansion, the isovelocity zones take the form of
cones with $\vz = -\vinf\mu$, so that constant \vz\ implies fixed
$\mu$.  The integral to be evaluated is then

\begin{equation}
\label{flform}
F_l(\vz) = \frac{2\pi}{D^2}\,\int_{r_{\rm min}}^\infty\,\Snu(r)
	e^{-\tau_{\rm c}(r,\mu)}\,r \, \sin^2\theta \, dr,
\end{equation}
 
\noindent where we have substituted for $p\,dp = \sin^2\theta\,r\,dr$,
and $r_{\rm min}$ equals $R_*$ for $\vz<0$, but $R_*/\sin\theta$ for
$\vz>0$ owing to stellar occultation.  An important point here is that
the integral is over radius, so that the angular dependence is a
constant of the integration.

\begin{figure}
\plotone{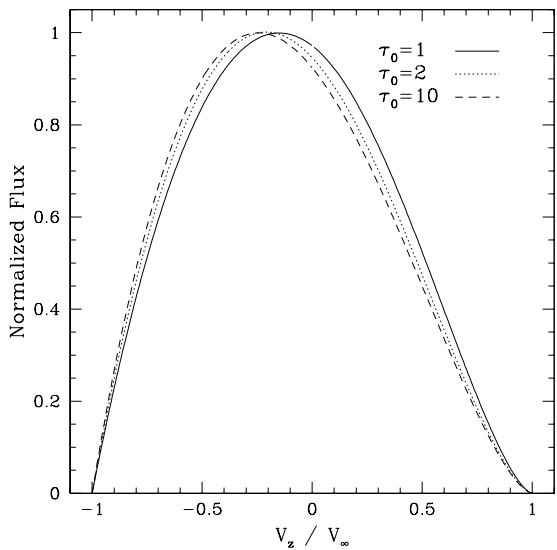}
\caption[]{A comparison of normalized thick line profiles showing the
influence of the wind attenuation with $\tau_0 = 1,$ 2, and 10 as
indicated.  As long as the line is optically thick in resonance line
scattering, the width and centroid shift of the line peak are not
strongly affected by the wind photoabsorption.

\label{fig:f3}}
\end{figure}

With a change of variable $u=R_*/r$, the integral reduces to

\begin{eqnarray}
\label{flsolve}
F_l(\vz) & = & \frac{2\pi\,R_*^2\,\epsilono\,\Bnu}{D^2}\,\left(\frac{3\,T}{2}
	\right)\,\sin^2\theta\,\nonumber \\
 & & \times\,\int_0^{u_{\rm max}}\,e^{-\tau_0 u\theta
	/\sin\theta}\,du. \\
 & = & \frac{3\pi\,R_*^2\,\epsilono\,\Bnu}{D^2}\,\nonumber \\
 & & \times\,\left(\frac{T}{\tau_0}
	\right)\,\left(\frac{\sin^3\theta}{\theta}\right)\,\left[ 
	1-e^{-\tau_0 u_{\rm max}\theta/\sin\theta} \right],
\end{eqnarray}

\noindent 
where $\theta = {\rm cos}^{-1}(-v_z/v_\infty)$, and
$u_{\rm max}=1$ for $\vz<0$ and $u_{\rm max}=\sin\theta$
for $\vz>0$.  Figure~\ref{fig:f3} shows normalized emission profiles at
various $\tau_0$.  The case of $\tau_0=1$ indicates that the wind
attenuation of X-rays is only mild, which invalidates our approximation
that that X-rays emerge only where $v_{\rm r} = \vinf$; however, the
$\tau_0=1$ case is included to demonstrate that the profile shape
varies qualitatively little for a broad range of $\tau_0$ values.

At large $\tau_0$, the line profile reduces to one of fixed shape 
described by

\begin{equation}
F_l(\vz) = \frac{3\pi\,R_*^2\,\epsilono\,\Bnu}{D^2}\,\left(\frac{T}{\tau_0}
        \right)\,\left(\frac{\sin^3\theta}{\theta}\right),
\end{equation}

\noindent 
which is to be contrasted against the analogous
asymptotic limit for thin lines from Ignace (2001):

\begin{equation}
F_l(\vz) = \frac{2\pi\,j_0\,R_*^3\,\lambda_0}{D^2\,\vinf}\,\left(\frac{1}
	{\tau_0}\right)\,\left(\frac{\sin\theta}{\theta}\right).
\end{equation}
 
\begin{figure}
\plotone{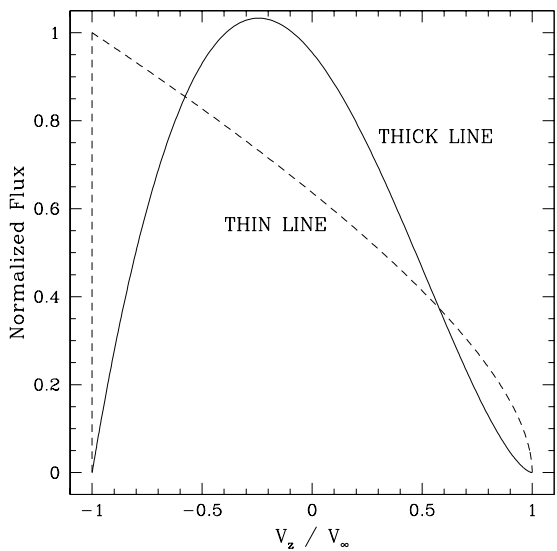}
\caption[]{A comparison of thin versus thick X-ray emission lines,
both in the asymptotic regime of large $\tau_0$.  Both lines
are asymmetric with a blueshifted centroid of the
emission peak, but the thick line is more symmetric than is the thin case.

\label{fig:f4}}
\end{figure}

\noindent The profile morphologies of the thin and thick asymptotic
forms are compared in Figure~\ref{fig:f4}.  Note that both are
asymmetric, with peak emission occuring at blueshifted velocities.
However, the thick line is much more nearly symmetric in appearance, and its
peak emission occurs not far from line center.  In fact for the
asymptotic limit of $F_l \propto \sin^3\theta/\theta$, the location
of the peak and the FWHM value of the profile can be easily computed.  Peak
emission occurs at $\vz = -0.24\times \vinf$, and the FWHM is
$1.26\times \vinf$ (alternatively, the HWHM is $0.63 \times \vinf$),
which holds for {\it all} thick lines with $T\gtrsim$ a few.  However,
although resonance line scattering does conserve photon number, photons
could be destroyed by other processes since there is a longer
``dwell'' time for photons than would be the case for optically thin
lines that scatter only once.  The next section considers the
influence of reabsorption effects for the net line emission and
profile shape.

\section{OPACITY EFFECTS ON ABUNDANCE DETERMINATIONS}

Optically thick resonance zones not only alter the profile shape,
they also modify the fraction of X-ray line photons that are absorbed
by the background continuum opacity prior to escape from the wind.
They do this in two distinct ways.  First, the preference for
azimuthal emission slightly alters the average column depth through
which the escaping flux must pass, and this changes the equivalent
width of the line profiles.  Second, if the Sobolev length $l_{\rm
sob}$ is not negligible, as would be the case for lines that are
highly broadened due to X-ray temperatures or microturbulence, then
optically thick resonance zones can ``trap'' photons long enough
to increase the total pathlength over which continuum absorption
can occur (Hillier 1983).  
Both of these processes were calculated accurately by
Hummer \& Rybicki (1985).  If these effects increase the reabsorption,
then observed line equivalent widths will be reduced, and anomalously
low abundances might be inferred.  Indeed, this is the interpretation
given by Kitamoto et al. (2000) to the substantial shortfall in
metal abundance inferred from optically thin fits to ASCA X-ray
line observations of several OB stars.  However, they did not apply
the theory of Hummer \& Rybicki (1985) in making their estimates,
and here we show that they overestimated the degree to which photon
destruction between scatterings in the line can reduce the observed
line flux.

\subsection{The Equivalant Width for Optically Thick Line Emission}

To quantify the change in absorption due to the first effect above,
note that optically thick lines in constant-expansion winds emit
photons preferentially in directions azimuthal to the flow.  Thus
optically thick line photons emitted in generally outward directions
(i.e., emitted toward the observer from the {\it near} side of the
wind) will have an overall reduced chance of escape, since more of
them are emitted from high impact parameters which experience
greater continuum optical depth for a given radius.  Conversely,
the photons observed from the {\it far} hemisphere will be augmented
on average, because then emitting more photons at higher impact
parameters implies less occultation by the underlying wind.  To
ascertain which of these competing influences dominates, we write
the expression for the ratio of the photon escape rate in the
optically thick and thin limits, which also gives the equivalent-width
($EW$) ratio

\beq \label{ewrat}
{EW_{\rm thick} \over EW_{\rm thin}} \ = \ {\int_{R_*}^\infty dr \ r^2 \rho^2
\int_{0}^{\pi} d\theta \ 1.5 \ \sin^3 \theta \ e^{-\tau_0R_*
\theta/r\sin{\theta}} \over \int_{R_*}^\infty dr \ r^2 \rho^2
\int_{0}^{\pi} d\theta \ e^{-\tau_0 R_* \theta/r \sin{\theta}}} \ .  
\eeq 

\noindent In the above expression, equation~(\ref{eq:tauc}) was
used for the photoabsorptive optical depth, and the normalized
emission profile from an optically thick resonance zone at constant
expansion speed is $1.5 \sin^2 \theta$.  It was also assumed that
the line emissivity scales as $\rho^2$, and that all line photons
eventually escape their natal resonance region.  Later we will
consider the fraction that may be absorbed within the resonance
zone.

Applying the condition $\rho \propto r^{-2}$ for constant expansion
allows the elementary radial integrals in equation~(\ref{ewrat})
to be carried out explicitly, giving in the limit of large total
radial wind optical depth $\tau_0$

\beq
{EW_{\rm thick} \over EW_{\rm thin}} \ = \ 1.5 \left(\int_0^\pi d\theta \ 
{\sin^4 \theta \over \theta}\right) \cdot \left ( \int_0^\pi d\theta \ {\sin^2 
\theta \over \theta} \right )^{-1} \ .
\eeq

\noindent When this expression is evaluated numerically, it is found
that
optically thick angular emission actually enhances escape, so the
far-hemisphere effect dominates.
This is the opposite sense as would be needed to help explain the
abundance anomalies, but is of little significance because
the magnitude of the increase in the equivalent
width is only 2\%.

\subsection{Reduction in Line Equivalent Width By Continuum Reabsorption}

The second effect mentioned above, reabsorption {\it within} the
finite region in which the X-ray line photons are created and
scattered, will certainly reduce the equivalent width when the
lines become optically thick.  This destruction mechanism is assumed
to be dominated by continuum absorption, rather than thermalization
within the line itself, and was considered by Kitamoto et al. (2000)
to be a promising solution to low equivalent-width anomalies in
ASCA observations of several OB winds.  Unfortunately, they assume
that photon escape from the X-ray-line resonance zone is diffusive,
which produces an overly large total pathlength during which
reabsorption can occur.  Instead, escape of photons from the
line-scattering region is known to occur by a ``single-flight''
escape process (Hummer \& Rybicki 1982), whereby Doppler redistribution
enables photons to rapidly reach the line wings.  At that point
the mean-free-path approaches the scale of the region and the escape
is non-diffusive.  In such a process, the total pathlength traversed
scales linearly with the length scale for escape, not quadratically
as assumed by Kitamoto et al. (2000).

Nevertheless, the total pathlength is indeed enhanced by optically
thick line scattering, so some of the abundance anomaly may be
explained by this process.  The basic theory of how continuum
opacity affects line formation in the Sobolev approximation was
developed by Hummer \& Rybicki (1985), so only a rough sketch of
their derivation is included here, expressed in intuitively motivated
probabilistic terms.  For a two-level atom in complete frequency
redistribution, common assumptions for line formation in the Sobolev
approximation, the line source function may be written

\beq
S_l \ = \ {\varepsilon B \over (\varepsilon \ + \ \xi)} \ ,
\eeq
where $\varepsilon$ is the destruction probability per scattering
and is assumed to be small, $B$ is the Planck function, and $\xi$
gives the fraction of radiative downward transitions in the atom
that are not balanced locally by radiative upward transitions.
Since $\varepsilon B$ here describes the photon creation rate, and
it is likely that $\varepsilon \ll \xi$, the essential statement
for our purposes is that $S_l \ \propto \ \xi^{-1}$, due to multiple
scattering within the optically thick resonance layer.  Hence
$\xi^{-1}$ may be associated with the expected number of scatterings
each photon undergoes between creation and escape.

In the first-order escape probability approximation, which is
especially applicable in the spirit of the Sobolev approximation,
$\xi$ is replaced by the probability that a line photon which has
just been emitted will {\it not} be scattered again in the line.
If the only other possibility is escape, then in the notation above
this implies $\xi \cong \beta$.  However, the presence of continuum
opacity allows for another alternative to scattering, which is
absorption within the resonance zone.  Thus $\xi$ is increased,
and $S_l$ reduced, by the presence of continuum absorption (e.g.,
Puls \& Hummer 1988).

The expression for the probability that no line scattering will
occur is most easily found by first writing the probability that
a line scattering {\it will} occur.  This is pieced together by
first noting that the photon is assumed to be emitted isotropically,
so is emitted randomly in a flat distribution over direction cosine
$\mu$.  In complete redistribution, it will also be emitted randomly
over the frequency profile $\phi(x)$, typically a Doppler profile.
Given an emission within $d\mu$ of angle $\mu$ and within $dx$ of
frequency $x$, in the Sobolev approximation in an expanding wind,
the photon propagates effectively not in real space but rather
toward redder comoving frequencies $x'$.

If the photon makes it to some $x'$ along direction $\mu$, the
probability of scattering within $dx'$ of $x'$ is given simply by
$\tau_\mu \phi(x') dx'$, where $\tau_\mu$ is the Sobolev optical
depth of the resonance zone along direction $\mu$.  In the absence
of continuum opacity, the probability that the photon will make it
to $x'$ is thus given by

\beq
P(x,x',\mu) \ = \ e^{-\tau_\mu \left [\Phi(x) - \Phi(x') \right ]} \ ,
\eeq
where
\beq
\Phi(x) \ - \ \Phi(x') \ = \ \int_{x'}^x dy \ \phi(y) \ ,
\eeq
and it is canonical to set $\Phi(-\infty) = 0$ via the definition
\beq
\Phi(x) \ = \ \int_{-\infty}^x dy \ \phi(y) \ .
\eeq
In all, one obtains
\beq
\label{rhonoc}
\xi \ \cong \ \beta \ = \ 1 \ - \ {1 \over 2} \int_{-1}^1 d\mu 
\int_{-\infty}^{\infty} dx \ \phi(x) \int_{-\infty}^{x} dx' \
\tau_\mu P(x,x',\mu) \ .
\eeq
Here $\tau_\mu$ and $\beta$ are given in equations (\ref{taulgen}) 
and (\ref{betaapp})
respectively, and if we denote the azimuthal Sobolev line optical
depth as
\beq
\tau_l \ = \ T \,{R_* \over r} \ ,
\eeq
then equation~(\ref{rhonoc}) yields, in agreement with equation~(\ref{betaapp}),
\beq
\xi \ \cong \ {2 \over 3 \tau_l} \ .
\eeq

The introduction of continuum opacity simply reduces $P(x,x',\mu)$
by an additional factor which represents the probability of continuum
absorption during propagation from $x$ to $x'$.  If the effective
line optical depth over the interval $dx'$ is $dx' \tau_\mu \phi(x')$,
then the continuum optical depth over that propagation interval is
$dx' \tau_\mu C$, where

\beq
C \ = \ {\kappa_{\rm c} \over \kappa_l}
\eeq
and $\kappa_{\rm c}$ and $\kappa_l$ are respectively the continuum
and mean line opacities.  By its definition, the continuum-absorption
parameter $C$ is readily shown to be related to the ratio of
$\tau_{\rm c}$, the radial continuum optical depth to the resonance
zone in question, and $\tau_l$, the azimuthal Sobolev optical depth
of that resonance zone, via

\beq
C \ = \ \left({\tau_0 \over T} \right) \left({v_{\rm th} \over v_{\infty}}\right)  \ = \
\left({\tau_{\rm c} \over \tau_l} \right) \left({v_{\rm th} \over v_{\infty}}\right)
\eeq
in a steady spherical wind at constant speed.

Note once again that here $v_{\rm th}$ represents all of the velocity
broadening over the Sobolev length, and so may include more than
the thermal speed of the ion in question and may approach the
hydrodynamical sound speed of the hot gas, which is at the level
of tens of percent of the terminal speed.  If $v_{\rm th}$ is much
greater than about 1/10 of $v_{\infty}$, the Sobolev approximation
itself may come into question, so this may be viewed as a rough
upper bound for the broadening to which our theory can apply.
Nevertheless, since $\tau_{\rm c} \sim 1$ where the observable line
emission forms, and $\tau_l \gae 10$ is required to apply optically
thick line theory, the condition $v_{\rm th} \lae v_{\infty}/10$
implies that $C \lae 10^{-2}$.  This unusually large upper bound
for the continuum opacity, relative to the lines, is a manifestation
of an implicit assumption that the hot line-forming gas represents
only a small fraction of a wind dominated by cooler and
continuously-absorbing gas.

By integrating the continuum optical depth encountered by a photon
propagating from $x$ to $x'$, the absorption-modified expression for
$P(x,x',\mu)$ then becomes
\beq
P(x,x',\mu,C) \ = \ e^{-\taul \left [\Phi(x)-\Phi(x')\right ]}
\times e^{-\taul C (x-x')} \ ,
\eeq
and so when $\xi$ is approximated
by the new probability that a scattering does {\it not} occur, the result is
\begin{eqnarray}
\Delta \xi(C) & = & 1 \ - \ {1\over 2} \int_{-1}^{1} d\mu 
\int_{-\infty}^{\infty} dx \ \phi(x)  \nonumber \\
 & & \times\,\int_{-\infty}^{x} dx'\phi(x') \tau_\mu P(x,x',\mu,C) \ .
\end{eqnarray}
This can be written
\beq
\xi(C) \ = \ \xi(0) \ + \ \Delta \xi(C)
\eeq
where
\beq
\xi(0) \ = \ \beta \ = \ {2 \over 3\tau_l}
\eeq
and
\begin{eqnarray}
\Delta \xi(C) & = & \int_0^1 d\mu \int_{-\infty}^{\infty} dx \
\phi(x) \nonumber \\
 & & \times\,\int_{-\infty}^{x} dx' \phi(x') \tau_\mu P(x,x',\mu,C)
\end{eqnarray}
was derived in an alternate notation by Hummer \& Rybicki (1985).

The correction to $\xi$ is found to become significant when $C\tau_l \sim 0.1$,
and when this is true, continuum reabsorption within the
line approaches the static limit for an infinite atmosphere.
Then the tabulated results of Hummer (1968)
show that a reasonable approximation for $\Delta \xi(C)$ when
$C$ is within an order of magnitude of $10^{-2}$ is
\beq
\Delta \xi(C) \ \cong \ 4 C \ .
\eeq
This implies that the factor by which the line source function $S_l$ must be
reduced to account for continuum absorption when
$C \tau_l \sim 0.1$ is
\beq
\label{srat}
{S_l(C) \over S_l(0)} \ \cong \ {\xi(0) \over \xi(C)} \ \cong \
{\beta \over (\beta + \Delta \xi(C))} \ \cong \ {1 \over (1 + 6 C \tau_l)} \ .
\eeq

\begin{figure}
\plotone{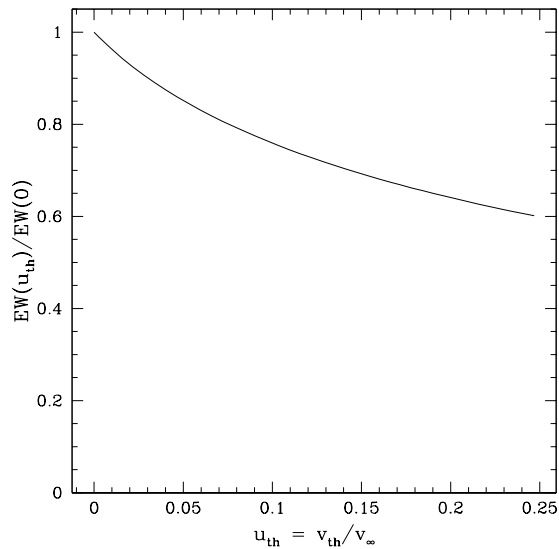}
\caption[]{A plot of how the equivalent width ($EW$) of thick lines
are reduced as a function of the degree of continuum reabsorption
as parametrized by the line broadening $u_{\rm th}= v_{\rm
th}/v_\infty$.  The normalization $EW(0)$ corresponds to the line
equivalent width with $u_{\rm th}=0$.  In the presence of
microturbulence, values of $u_{\rm th}$ may rise to a couple of tenths
leading to a modest reduction (20--40\%) in the line $EW$.

\label{fig:f5}}
\end{figure}

As mentioned above, $C \tau_l \sim v_{\rm th}/v_{\infty}$ near the
X-ray photosphere, so when $v_\infty$ is of order $10^3$~km~s$^{-1}$,
$C \tau_l$ can reach values as large as 0.1 only when the line
broadening represents a substantial fraction of the sound speed in
the X-ray gas.  Interestingly, although the parameter $C$ does
affect the source-function correction at any given radius, it is
not directly relevant to the integrated equivalent width, because
increasing $\kappa_{\rm c}$ simply moves the resonance zones of
interest out to larger radii and therefore lower density.  This
regulates the reabsorption within the resonance zone in a manner
that is independent of $C$.  Instead, it is the parameter $v_{\rm
th}/v_\infty$ which controls the correction to the equivalent width,
so we give this parameter its own designation, $u_{\rm th}$, and
note that $C \tau_l = u_{\rm th} \tau_{\rm c}$.  The equivalent
width correction is calculated by inserting $S_l(C)/S_l(0) =
(1+\Delta \xi/\beta)^{-1}$ from equation~(\ref{srat}) into the
integral in the numerator of equation~(\ref{ewrat}), and the result
is shown in Figure~\ref{fig:f5}.

It is apparent from this figure that the equivalent-width correction
is only substantial when $u_{\rm th}$ is appreciably elevated beyond
the purely thermal motions of the heavy ions.  This might occur,
for example, if overpressures at the sonic scale in the hot gas
were to accelerate flows stochastically on the scale of the Sobolev
length, which would yield superthermal microturbulence in the X-ray
lines.  In such a scenario, Figure~\ref{fig:f5} shows that the observed
equivalent widths could be reduced by up to about 40\% before the
basic assumptions of this model are severely violated.  Such a
substantial reduction could certainly alter the inferred abundance
of the ion involved, although perhaps not to the degree expected
by Kitamoto et al. (2000).  Whether or not it could provide a
complete explanation for the observed abundance anomalies remains
a question for more detailed physical models; the sole conclusion
here is that if abundance anomalies are to be explained by continuum
absorption within the line resonance zones, some type of microturbulent
broadening is required to generate Sobolev lengths with $l_{\rm
sob} \gae 0.1 r$, straining the Sobolev approximation itself.  This
possibility is not beyond reason, however, as a radial variant on
this same concept has already been invoked (Lucy 1982) to explain
``black troughs'' in UV resonance-line spectra of OB winds (Lamers
\& Morton 1976; Prinja, Barlow, \& Howarth 1990).


\subsection{The Influence of Continuum Absorption on Line Profile Shape}

Appreciable values of $C\,\tau_l$ not only effect the equivalent
width in low-resolution spectra, they also alter the profile shape
at high resolution.  Since $\tau_{\rm c} = \tau_0 (R_*/r)$ implies that
$C\tau_l \propto r^{-1}$, the effect of continuum absorption on
the line becomes less important at large radii.  Thus when the
factor given by equation~(\ref{srat}) is inserted into equation
(\ref{flform}), the profile shape is altered by the fact that $S_l$
will now fall less rapidly than the $r^{-3}$ of equation~(\ref{snur}).
Since $C$ is fairly small, this will only be important when $\tau_l$
is large, and in that limit the modified equation~(\ref{flsolve})
becomes

\beq
\label{ewratreab}
F_l(\vz) \ = \ 
\frac{3\pi\,R_*^2\,\epsilono\,\Bnu}{D^2}\,\left(\frac{T}{\tau_0}
        \right)\,\left(\frac{\sin^3\theta}{\theta}\right)\,
f \left (
{1 \over 6}{v_\infty \over v_{\rm th}} {\theta \over {\rm sin} \theta} \right ) \ ,
\eeq
where we have defined
\beq
f(x) \ = \ x e^x \int_x^\infty dy \ {e^{-y} \over y} \ .
\eeq

\begin{figure}
\plotone{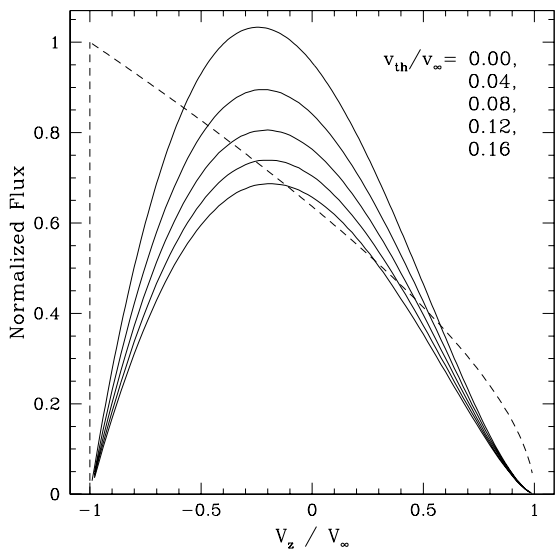}
\caption[]{A plot of different thick line profiles as $v_{\rm
th}/v_\infty$ is increased from 0 to 0.16 (solid lines), with a
thin line profile shown for comparison (dashed).  The influence of
reabsorption is twofold:  larger values of the broadening diminish
the overall amount of line emission, and the profile becomes more
symmetric in shape.  In the unphysical limit of $v_{\rm th}/v_\infty
\gg 1$, the profile will take on the purely symmetric
form of $(1-\vz^2/\vinf^2)$.

\label{fig:f6}}
\end{figure}

This profile shape is plotted in Figure~\ref{fig:f6} for a range
in values of $v_{\rm th}/v_{\infty}$.  Note that if $v_{\rm th}
\ll v_\infty$, $f$ approaches unity and we recover the $\tau_0 \gg
1$ limit of equation~(\ref{flsolve}).  The unphysical limit $v_{\rm
th} \gg v_\infty$ can be used as a check, since then $f(x)$ is of
order $x$, and the profile approximates the (symmetric) parabolic
shape $1-v_{\rm z}^2/v_\infty^2$ characteristic of $\mu$-dependent
line emissivity over a ``skin depth'' of fixed continuum mean-free-path.
This limit is of course unphysical because it completely violates
the approximations made, particularly the Sobolev approximation.
Instead, the importance of the profile shapes seen in Figure~\ref{fig:f6}
applies for much more modest values of $v_{\rm th}/v_\infty$, which
are seen to reduce the profile asymmetry even further than that
due solely to continuum reabsorption {\it outside} the resonance
zone.

\section{SUMMARY AND DISCUSSION} 	\label{sec:disc}

Our paper considers the role of optical depth in X-ray emission-line
formation in stellar winds.  The line photons are assumed to be created
by collisional processes, and thus scale with the square of the
density, and may then be resonantly scattered.  We have demonstrated
that strong lines with sufficient opacity to multiply scatter emerging
photons will alter the observed profile shape in a recognizably
characteristic way.  Our line synthesis has taken into account the
``cool'' wind photoabsorption, which can strongly attenuate X-ray
photons at soft and intermediate energies.

We argue that when the wind is highly optically thick to
photoabsorption, it is adequate to consider constant expansion for the
purpose of computing the line profile.  In this limit the expressions
for the radiative transfer can be solved analytically.  Lines that are
optically thick over a resonance zone are found to have profile shapes
that scale as $F_l \propto \sin^3 \theta / \theta$, which is to be
compared to the analogous thin-line case of $\sin \theta / \theta$,
where the observed velocity shift is $\vz = - \vinf\, \cos \theta$.

\begin{figure}
\plotone{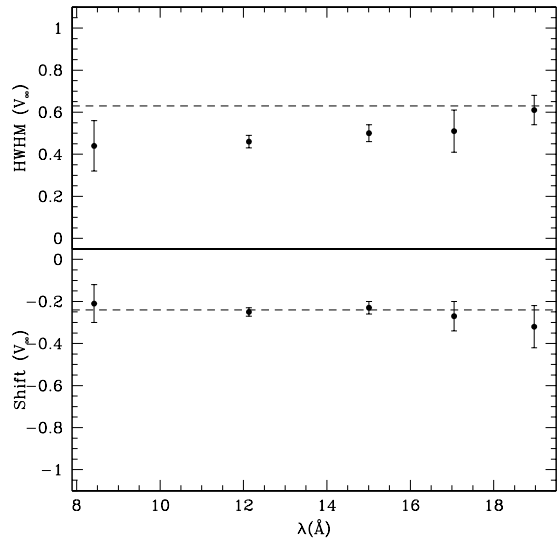}
\caption[]{A comparison of the HWHM and centroid shift (both
normalized by \vinf) for $\zeta$~Pup observed with {\it Chandra}
versus the predictions of our limiting thick line case.  The points
correspond to five different lines with errorbars as tabulated in
Cassinelli \etal\ (2001).  Although the HWHMs are slightly smaller
than predicted, the centroid values agree well with our theoretical
results.  Curiously, lines at longer wavelength appear more nearly
consistent with our predicted result.  At the longer wavelengths,
the line emission escapes from larger radius (see Fig.~\ref{fig:f1})
where the constant expansion approximation that we have used is
best satisfied.

\label{fig:f7}}
\end{figure}

It useful to compare these theoretical values against those measured
for $\zeta$~Pup by Cassinelli \etal\ (2001).  These authors list the
HWHM and centroid shift of five lines in their Table~2 (we ignore the
\ion{N}{7} profile which is clearly different owing to its flat-topped
morphology).  The centroid shifts range between $-460$ and $-700$
\kmsec, and the HWHMs range from 970 to 1340~\kmsec.  Cassinelli
\etal\ quote a terminal speed of 2250~\kmsec.  Hence peak line
emissions fall between $-0.2$ and $-0.3$ of \vinf, and the HWHM
are around 0.4 to 0.6 of \vinf.  This data is compared with the
asymptotic thick line predictions in Figure~\ref{fig:f7} (i.e.,
without reabsorption effects).  The HWHMs all fall below our
predicted value, but the centroid values are all consistent with
our thick line results.  This is somewhat surprising in light of
our rather simplified assumptions and the fact that we do not allow
for any radial variation of the hot gas temperature.  We must
interject the caution that we have not actually tried to {\it fit}
the observed profiles, only compare global characterizations in
the form of where peak emission occurs and the overall line width.
More detailed line fitting to the data is necessary to determine
categorically whether thin or thick lines better represent each
observation, but our simple comparison suggests that optical-depth
effects could be a key element to interpreting the observations,
and that the $\zeta$~Pup data is more consistent with the standard
wind-shock paradigm than previously thought.

However for $\zeta$~Ori, $\theta^1$ Ori~C, and $\delta$~Ori, the
observed emission profiles are shown or claimed to be extremely
symmetric, with little or no centroid shift.  (Note that for observed
line profiles of modest S/N and spectral resolution, our theoretical
line shapes could match the symmetric appearance only if the central
line wavelengths were misplaced slightly to the blue.) Thick
resonance-line scattering by itself is not able to explain such
symmetries, at least not without allowing parameters such as $T_X$
to vary with radius, or introducing new free parameters.  However,
a consideration of reabsorption by the continuum opacity within a
highly optically thick resonance zone can in principle lead to
symmetric lines with an inverted parabolic shape in the limiting
case, and suppression of the line emission.  The effect depends on
the relative line broadening $v_{\rm th}/v_\infty$, leading to a
modest effect for only thermal broadening by the hot gas, but can
significantly reduce the line equivalent width ($\approx 40\%$) if
there are extreme microturbulent motions of order $0.25 v_\infty$.
For this latter case, the influence of reabsorption by thick
resonance-line scattering may help to understand both the line
symmetry in these three stars and the sub-solar abundance problem.
One major difficulty is whether observed lines have sufficiently
high values of $T$ so that our results apply.  A second is, where
are the optically thin lines that do show the theoretically expected
morphology for distributed wind shocks, and do such lines also
exhibit abundance problems?

One effect that we have ignored has been the influence of the wind
acceleration zone.  In lower-density winds, the formation of lines
at intermediate X-ray energies will sample these inner radial depths
of the wind.  This will lead to a source function of the form $\Snu
\propto (\rho_{\rm w}/\beta) = \tau_l/(r^2\vrad)$, where $\vrad$
varies rapidly in the span of just 1--2$R_*$ and $\tau_l$ depends
on the velocity gradient.  However, the wind attenuation will also
be enhanced, plus reabsorption effects of line photons will be more
severe.  So there are strongly competing effects between shifting
the profile to greater symmetry versus asymmetry, and a more detailed
analysis will be required in general.

Another effect that has been ignored is the influence of ``porosity''.
We have treated the hot X-ray emitting gas as being thoroughly mixed
throughout the ambient wind flow.  However, the wind is probably highly
structured in density and velocity, both radially and laterally.  If
the porosity is severe, in the sense that cool absorbing wind
material is strongly clumped such that the mean-free-path for X-ray
photons becomes quite long, then this may explain the lack of
profile symmetry in $\zeta$~Ori, $\theta^1$~Ori~C, and $\delta$~Ori
(Cassinelli 2001, private communication).  Of course, one must then
pose the question of why $\zeta$~Pup is different.

Finally, should optical-depth effects be relevant for other early-type
stars?  The B~star winds are far less dense than O~stars, so that
except at the softest energies, the X-rays from the entire wind
probably emerge free of attenuation.  In this case, the line synthesis
presented by Owocki \& Cohen (2001) will likely be able to match the
profiles, should they be observed.  On the other hand, the Wolf-Rayet
winds are substantially denser than for the O~stars, so dense that
$r_1$ is 10's and even 100's of $R_*$ in extent.  In this case, strong
lines are probably optically thick over much of the wind, but this is
compensated by the fact that the observed line emission emerges from
large radius.  Such lines can certainly be assumed to form in a
constant-expansion wind, so if thin, the asymptotic profile shape
derived by Ignace (2001) should apply, and if thick, the asymptotic
profile shape derived here applies instead.  These conclusions need to
be tested by future observations.

\acknowledgements

We express appreciation to Joe Cassinelli, Stan Owocki, John Hillier,
and David Cohen for comments in relation to the possibility of
thick X-ray lines.  We also gratefully acknowledge the referee,
John Castor, for several insightful and helpful remarks.  This
research was supported by NASA grant NAG5-9964.

\end{document}